\documentclass{article}
\usepackage{amsfonts}
\usepackage{amsmath}
\usepackage{graphicx}
\usepackage{cite}

\begin{document}

\title{Van der Pol - Duffing oscillator: global view of metamorphoses of the
amplitude profiles}
\author{Jan Kyzio\l , Andrzej Okni\'nski \\
%EndAName
Politechnika \'Swi\c{e}tokrzyska, Al. 1000-lecia PP 7, \\
25-314 Kielce, Poland}
\maketitle

\begin{abstract}
Dynamics of the Duffing--Van der Pol driven oscillator is investigated.
Periodic steady-state solutions of the corresponding equation are computed
within the Krylov-Bogoliubov-Mitropolsky approach to yield dependence of
amplitude $A$ on forcing frequency $\Omega $ as an implicit function, $%
F\left( A,\Omega \right) =0$, referred to as resonance curve or amplitude
profile.

In singular points of the amplitude curve the conditions $\frac{\partial F}{%
\partial A}=0$, $\frac{\partial F}{\partial \Omega }=0$ are fulfilled, i.e.
in such points neither of the functions $A=f\left( \Omega \right) $, $\Omega
=g\left( A\right) $, continuous with continuous first derivative, exists.
Near such points metamorphoses of the dynamics can occur. In the present
work the bifurcation set, i.e. the set in the parameter space, such that
every point in this set corresponds to a singular point of the amplitude
profile, is computed.

Several examples of singular points and the corresponding metamorphoses of
dynamics are presented.
\end{abstract}

\section{Introduction}

\label{intro}

In this paper we continue our study of the Duffing - van der Pol (DvdP)
oscillator \cite{Kyziol2015a} which has been extensively investigated due to
exhibiting interesting, complicated dynamics and to potential applications
in physics, chemistry, biology, engineering, electronics, and many other
fields, see \cite%
{Szemplinska1997,Chudzik2011,Brezetskyi2015,Vincent2015,Wiggers2018,Bi2004,Chen2010,YueYu2017,Nijah2008,Luo2014}%
.

In \cite{Kyziol2015a} we studied metamorphoses of the amplitude profiles,
obtained within the Krylov-Bogoliubov-Mitropolsky (KBM) approach, induced by
changes of control parameters near singular points of these curves. In the
present work we determine the bifurcation set -- the space of parameters $C$%
, such that for every $C$ the amplitude curve $L\left( \Omega ,\ A;\
C\right) =0$ has a singular point (this definition bears some similarity to
the definition of the bifurcation set in the Catastrophe Theory \cite%
{PostonStewart2014}). Since qualitative changes of dynamics occur in
neighbourhoods of singular points \cite{Kyziol2013,Kyziol2015b,Kyziol2017}
we obtain in this way a global view of bifurcation parameters.

We consider the more general periodically forced Duffing - van der Pol
oscillator with nonlinear damping which can be written as:%
\begin{equation}
\dfrac{d^{2}x}{dt^{2}}-\left( b-cx^{2}\right) \dfrac{dx}{dt}+e\left( \frac{dx%
}{dt}\right) ^{3}+ax+dx^{3}=f\cos \omega t.  \label{DvdPnd1}
\end{equation}

There are three main cases of the Duffing potential $V\left( x\right) =\frac{%
1}{2}ax^{2}+\frac{1}{4}dx^{4}$: (i) single well $\left( a>0\text{,}\
d>0\right) $, (ii) double well $\left( a<0\text{, }d>0\right) $, and (iii)
double hump $\left( a>0\text{,}\ d<0\right) $ . In the present paper we
consider cases (i), ( iii) and it is thus assumed that $a,b,c,f,\omega >0$
while $d,e$ are arbitrary. For $e=0$ the DvdP model is recovered.

We compute periodic steady-state solutions of the DvdP equation within the
Krylov-Bogoliubov-Mitropolsky (KBM) approach \cite{Nayfeh2011} to get
dependence of amplitude on forcing frequency as an implicit function,
referred to as resonance curve or amplitude profile. We investigate
metamorphoses of the computed amplitude profiles induced by changes of
control parameters near singular points of these curves since qualitative
changes of dynamics occur in neighbourhoods of singular points, see \cite%
{Kyziol2017} and references therein. The idea to use Implicit Function
Theorem in this context was proposed in \cite{Awrejcewicz1995}.

The paper is organized as follows. In the next Section the DvdP equation is
written in nondimensional form and implicit equation for resonance curves $%
L\left( \Omega ,\ A\right) =0$ is derived via the KBM approach. In Section %
\ref{general} theory of algebraic curves is applied to compute the
bifurcation set for the DvdP equation ($e=0$) which yields a global view of
bifurcations. In Section \ref{Computations} examples of metamorphoses of
dynamics occurring in neighbourhoods of singular points are presented. We
summarize our results in the last Section.

\section{Nonlinear resonances via KBM method}

\label{KBM}

A very exact generalized harmonic balance method was developed by Luo \cite%
{Luo2012}. We apply to Eq. (\ref{DvdPnd1}) less exact but sufficient in case
of steady-states the Krylov-Bogoliubov-Mitropolsky perturbation
approach \cite{Nayfeh2011}. Substituting into (\ref{DvdPnd1}): 
\begin{equation}
x=\sqrt{\tfrac{b}{c}}z,\ t=\tfrac{1}{\sqrt{a}}\tau ,\ \omega =\sqrt{a}\Omega
,  \label{defs}
\end{equation}%
we get the DvdP equation in nondimensional form:%
\begin{equation}
\begin{array}{c}
\dfrac{d^{2}z}{d\tau ^{2}}-\mu \left( 1-z^{2}\right) \dfrac{dz}{d\tau }+\nu
\left( \frac{dz}{d\tau }\right) ^{3}+z+\lambda z^{3}=G\cos \left( \Omega
\tau \right) , \\ 
\mu ,\ G,\ \Omega >0,\quad \lambda ,\ \nu \text{ -- arbitrary,}%
\end{array}
\label{DvdP2}
\end{equation}%
where $\mu =\frac{b}{\sqrt{a}}$, $\nu =\frac{\sqrt{a}be}{c}$, $\lambda =%
\frac{bd}{ac}$, $G=\frac{f}{a}\sqrt{\frac{c}{b}}$ \bigskip 

The equation (\ref{DvdP2}) is written in the following form: 
\begin{equation}
\frac{d^{2}z}{d\tau ^{2}}+\Omega ^{2}z+\varepsilon \left( \sigma z+g\right)
=0,  \label{DvdP3}
\end{equation}%
where 
\begin{equation}
\left. 
\begin{array}{rl}
g= & -\Theta _{0}^{2}z-\mu _{0}\dfrac{dz}{d\tau }+\mu _{0}z^{2}\dfrac{dz}{%
d\tau }+\nu _{0}\left( \frac{dz}{d\tau }\right) ^{3}+\alpha _{0}z+\lambda
_{0}z^{3}-G_{0}\cos \left( \Omega \tau \right) \\ 
\Theta _{0}^{2}= & \dfrac{\Theta ^{2}}{\varepsilon },\ \mu _{0}=\dfrac{\mu }{%
\varepsilon },\ \nu _{0}=\dfrac{\nu }{\varepsilon },\ \alpha _{0}=\dfrac{1}{%
\varepsilon },\ \lambda _{0}=\dfrac{\lambda }{\varepsilon },\ G_{0}=\dfrac{G%
}{\varepsilon },\ \varepsilon \sigma =\Theta ^{2}-\Omega ^{2}%
\end{array}%
\right\} .  \label{defg}
\end{equation}

According to the KBM method we assume for small nonzero $\varepsilon $\ that
the solution for $1:1$ resonance can be written as: 
\begin{equation}
z=A\left( \tau \right) \cos \left( \Omega \tau +\varphi \left( \tau \right)
\right) +\varepsilon z_{1}\left( A,\varphi ,\tau \right) +\ldots
\label{solz}
\end{equation}%
with slowly varying amplitude and phase:%
\begin{eqnarray}
\dfrac{dA}{d\tau } &=&\varepsilon M_{1}\left( A,\varphi \right) +\ldots \ ,
\label{solA} \\
\dfrac{d\varphi }{d\tau } &=&\varepsilon N_{1}\left( A,\varphi \right)
+\ldots \ .  \label{solphi}
\end{eqnarray}

Computing now derivatives of $z$ from Eqs.(\ref{solz}), (\ref{solA}), (\ref%
{solphi}) and substituting to Eqs.(\ref{DvdP3}), (\ref{defg}), eliminating
secular terms and demanding $M_{1}=0$, $N_{1}=0$ we obtain the following
equations for the amplitude and phase of steady states:

\begin{equation}
A^{2}\Omega ^{2}\left( \mu \left( 1-\tfrac{1}{4}A^{2}\right) -\tfrac{3}{4}%
\nu A^{2}\Omega ^{2}\right) ^{2}+A^{2}\left( 1+\tfrac{3}{4}\lambda
A^{2}-\Omega ^{2}\right) ^{2}=G^{2}.  \label{AOmega}
\end{equation}

\section{General properties of the amplitude profile $A\left( \Omega \right) 
$}

\label{general}

\subsection{Singular points}

After introducing new variables, $\Omega ^{2}=X$, $A^{2}=Y$, the equation ( %
\ref{AOmega}) defining the amplitude profile reads $L\left( X,Y; \lambda,
\mu, \nu, G\right) =0$ where: 
\begin{equation}
L\left( X,Y \right) =XY\left( \mu \left( 1-\tfrac{1}{4}Y\right) -\tfrac{3}{4}%
\nu YX\right) ^{2}+Y\left( 1+\tfrac{3}{4}\lambda Y-X\right) ^{2}-G^{2}=0.
\label{LXY}
\end{equation}
Singular points of $L\left( X,Y\right) $ are computed from equations \cite%
{Wall2004}: 
\begin{subequations}
\label{SING1}
\begin{eqnarray}
L &=&0,  \label{Sing1a} \\
\tfrac{\partial L}{\partial X} &=&0,  \label{Sing1b} \\
\tfrac{\partial L}{\partial Y} &=&0.  \label{Sing1c}
\end{eqnarray}
In \cite{Kyziol2015a} we have found several analytic solutions of Eqs. (\ref%
{SING1}) for the DvdP equation, $\nu =0$.

\subsection{Bifurcation set}

It follows from general theory of implicit functions that in a singular
point there are multiple solutions of equation (\ref{LXY}). We shall use
this property to compute parameters values for which singular points occur
for the DvdP equation (i.e. for $\nu =0$).

Equation (\ref{Sing1b}) means that the function $X=f\left( Y\right) $ does
not exist in the neighbourhood of a solution of Eqs. (\ref{Sing1a}), while
due to Eq. (\ref{Sing1c}) also the function $Y=g\left( X\right) $ does not
exist (more exactly, there are no continuous functions $f$, $g$ with
continuous first derivatives $\frac{df}{dY}$, $\frac{dg}{dX}$).

On the other hand, to define a singular point we can use equations ( \ref%
{Sing1a}), (\ref{Sing1b}) and an alternative of condition (\ref{Sing1c})
which excludes existence of the single-valued function $Y=g\left( X\right) $%
. We thus solve Eqs. (\ref{Sing1a}), (\ref{Sing1b}) obtaining equation for a
function $Y=g\left( X\right) $ and then demand that there are multiple
solutions for such $Y$ (alternatively, we could have solved Eqs. (\ref%
{Sing1a}), (\ref{Sing1c})).

Solution of Eqs. (\ref{Sing1a}), (\ref{Sing1b}), with $L\left( X,Y\right) $
given by Eq. (\ref{LXY}) (note that we have set $\nu =0$) can be written as
a quintic equation for $Y$: 
\end{subequations}
\begin{subequations}
\begin{equation}
\left. 
\begin{array}{l}
F\left( Y\right)
=a_{5}Y^{5}+a_{4}Y^{4}+a_{3}Y^{3}+a_{2}Y^{2}+a_{1}Y+a_{0}=0,\medskip \\ 
a_{5}=\mu ^{4},\ a_{4}=-16\mu ^{2}\left( \mu ^{2}+3\lambda \right) ,\
a_{3}=32\mu ^{2}\left( 3\mu ^{2}+12\lambda -2\right) ,\smallskip \\ 
a_{2}=-256\mu ^{2}\left( \mu ^{2}+3\lambda -2\right) ,\ a_{1}=256\mu
^{2}\left( \mu ^{2}-4\right) ,\ a_{0}=1024G^{2},%
\end{array}%
\right\}  \label{Sol1}
\end{equation}%
with $X$ given by: 
\begin{equation}
X=-\tfrac{1}{32}\mu ^{2}Y^{2}+\left( \tfrac{1}{4}\mu ^{2}+\tfrac{3}{4}%
\lambda \right) Y-\tfrac{1}{2}\mu ^{2}+1,  \label{Sol2}
\end{equation}
and we assume that $\mu \neq 0$.

Equation (\ref{Sol2}) is identical with Eq. (16) in (\cite{Kyziol2015a})
which for $G\neq 0$\ determines the general solution of Eqs. (\ref{SING1})
(note that the text after Eq. (14) in (\cite{Kyziol2015a}) should read:
equation for $\mu $: $B_{6}\mu ^{6}+B_{4}\mu ^{4}+B_{2}\mu ^{2}+B_{0}=0$).

Necessary condition for a polynomial to have multiple roots is that its
discriminant vanishes \cite{Gelfand1994}. Discriminant $\Delta $ can be
computed as a resultant of a polynomial $F\left( Y\right) $ and its
derivative $F^{\prime }$, with a suitable normalizing factor. Resultant of a
quintic polynomial $F\left( Y\right)
=a_{5}Y^{5}+a_{4}Y^{4}+a_{3}Y^{3}+a_{2}Y^{2}+a_{1}Y+a_{0}$ and its
derivative $F^{\prime }$ is given by determinant of the Sylvester matrix $S$%
, $\Delta =a_{5}^{-1}\det \left( S\right) $ where \cite{Gelfand1994}: 
\end{subequations}
\begin{equation}
S=\left( 
\begin{array}{ccccccccc}
a_{0} & a_{1} & a_{2} & a_{3} & a_{4} & a_{5} & 0 & 0 & 0 \\ 
0 & a_{0} & a_{1} & a_{2} & a_{3} & a_{4} & a_{5} & 0 & 0 \\ 
0 & 0 & a_{0} & a_{1} & a_{2} & a_{3} & a_{4} & a_{5} & 0 \\ 
0 & 0 & 0 & a_{0} & a_{1} & a_{2} & a_{3} & a_{4} & a_{5} \\ 
a_{1} & 2a_{2} & 3a_{3} & 4a_{4} & 5a_{5} & 0 & 0 & 0 & 0 \\ 
0 & a_{1} & 2a_{2} & 3a_{3} & 4a_{4} & 5a_{5} & 0 & 0 & 0 \\ 
0 & 0 & a_{1} & 2a_{2} & 3a_{3} & 4a_{4} & 5a_{5} & 0 & 0 \\ 
0 & 0 & 0 & a_{1} & 2a_{2} & 3a_{3} & 4a_{4} & 5a_{5} & 0 \\ 
0 & 0 & 0 & 0 & a_{1} & 2a_{2} & 3a_{3} & 4a_{4} & 5a_{5}%
\end{array}%
\right) .  \label{S}
\end{equation}
For explicit formula for a discriminant of the quintic polynomial see Eq.
(1.36) in Chapter 12 in \cite{Gelfand1994}.

\begin{figure}[ht!]
\center 
\includegraphics[width=12cm, height=8cm]{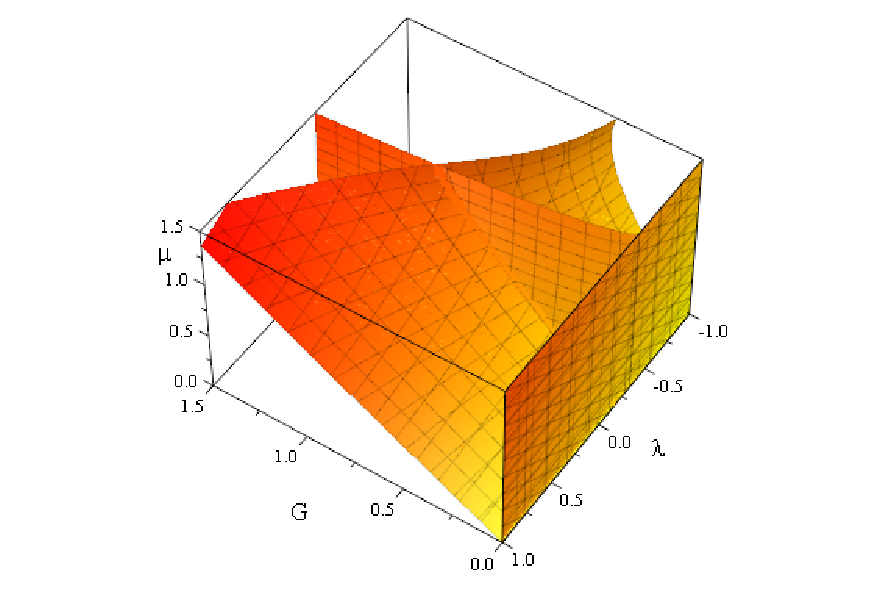}
\caption{The bifurcation set: $D\left( G,\protect\lambda ,\protect\mu %
\right) =0$, $G=0$, see Eq. (\protect\ref{Delta1})}
\label{F1}
\end{figure}

The discriminant $\Delta $ for polynomial $F(Y)$ in Eq. (\ref{Sol1}) is: 
\begin{equation}
\left. 
\begin{array}{l}
\Delta =2^{40}\mu ^{10}G^{2}\,D\left( G,\lambda ,\mu \right) ,\quad \left(
\mu \neq 0\right) \\ 
D\left( G,\lambda ,\mu \right) =A_{10}\mu ^{10}+A_{8}\mu ^{8}+A_{6}\mu
^{4}+A_{4}\mu ^{4}+A_{2}\mu ^{2}+A_{0},%
\end{array}%
\right\}  \label{Delta1}
\end{equation}%
with coefficients $A_{i}$ given in the appendix.

Therefore, condition $\Delta =0$ yields the bifurcation set: 
\begin{equation}
D\left( G,\lambda ,\mu \right) =0,\text{ or }G=0,  \label{Delta2}
\end{equation}
the manifold of singular points in the parameter space $\left( G,\lambda
,\mu \right) $, see Fig. 1 where a fragment of the surface $D\left(
G,\lambda ,\mu \right) =0$ as well as the plane $G=0$ are plotted:

The surface $D\left( G,\lambda ,\mu \right) =0$ may be considered as
consisting of two intersecting surfaces. There are thus three families of
singular points: lying on one surface, lying on another surface, and the
third family consists of points lying on a curve -- the intersection of
these two surfaces.

\subsection{Double singular points}

Double singular points -- points belonging to the self-intersection of the
surface $D\left( G,\lambda ,\mu \right) =0$ -- can be computed numerically.
It follows from Fig. \ref{F1} that for $\mu =\mu _{0}>0$ there is a singular
point on the curve obtained by intersecting the surface $D=0$ by the plane $%
\mu =\mu _{0}$, i.e. on the curve $D\left( G,\lambda ,\mu _{0}\right) =0$.
Therefore, the intersection curve can be computed in the following way. If
we set $\mu =\mu _{0}$, then $\lambda $, $G$ can be computed from equations: 
\begin{subequations}
\label{IS}
\begin{eqnarray}
\frac{\partial D\left( G,\lambda ,\mu _{0}\right) }{\partial G} &=&0,
\label{is1} \\
\frac{\partial D\left( G,\lambda ,\mu _{0}\right) }{\partial \lambda } &=&0,
\label{is2}
\end{eqnarray}%
which define a singular point on the curve $D\left( G,\lambda ,\mu
_{0}\right) =0$. Since there are several solutions of equations (\ref{IS}),
mostly complex, we have to choose the right solution $G_{0},\lambda _{0},\mu
_{0}$, i.e. such that fulfills $D\left( G_{0},\lambda _{0},\mu _{0}\right)
=0 $, i.e. lies on the surface $D=0$. It follows from Fig. \ref{F1} that for 
$\mu _{0}>0$ such solution exists.

There are also other double singular points -- belonging to intersection of
surfaces $\Delta =0$ and $G=0$. Solving equations: 
\end{subequations}
\begin{subequations}
\label{DELTAG}
\begin{eqnarray}
\Delta &=&0,  \label{G1} \\
G &=&0,  \label{G2}
\end{eqnarray}%
we obtain a very simple condition: 
\end{subequations}
\begin{equation}
4096\mu ^{4}\left( \mu -2\right) ^{2}\left( \mu +2\right) ^{2}\left(
3\lambda +1\right) ^{3}\left( 9\lambda ^{2}+\mu ^{2}+3\lambda \mu
^{2}\right) =0.  \label{isDeltaG}
\end{equation}

We shall investigate below the solution $\mu =2$, $\lambda $ -- arbitrary.
More exactly, we shall consider the case $\mu =2$, $\lambda \geq -\left\vert
a\right\vert $ with small values of $\left\vert a\right\vert $ since for
more negative values of $\lambda $ the system is unstable -- trajectories
escape to infinity.

\section{Computational results}

\label{Computations}

A global picture of singular points, shown in the bifurcation set, Fig. 1,
can be used to predict metamorphoses of dynamics. In our previous work \cite%
{Kyziol2015a} we have described metamorphoses occuring near singular poitns
belonging to the plane $G=0$. It is now obvious, that all points of the
plane $G=0$ in the parameter space $\left( G\text{, }\lambda \text{, }\mu
\right) $ are singular -- they are isolated points of the amplitude profile,
cf. Eq. (\ref{Delta2}). Such metamorphosis is described in Section 4.1 in 
\cite{Kyziol2015a} for $\lambda =1$, $\mu =0.5$.and $G\geq 0$.

In this work we study metamorphoses occuring near double singular points:
belonging to self-intersection of the surface $D\left( G,\lambda ,\mu
\right) =0$ and belonging to intersection of the surface $D\left( G,\lambda
,\mu \right) =0$ and the plane $G=0$, see Fig. 1.

We first compute a point lying on the self-intersection curve in Fig. 1. For 
$\mu _{0}=3$ we get from Eqs. (\ref{IS}) solution $\lambda _{0}=-0.118\,321$%
, $G_{0}=1.\,214\,842$ which fulfills $\Delta \left( G_{0},\lambda _{0},\mu
_{0}\right) =0$. Now we plot the corresponding amplitude profile, i.e. the
implicit function $A\left( \Omega \right) $ given by Eq. (\ref{AOmega}) for
parameters values $\lambda =\lambda _{0}$, $\mu =\mu _{0}$, $G<G_{0}$, $%
G=G_{0}$, $G>G_{0}$. The amplitude function $A\left( \Omega ;G_{0},\lambda
_{0},\mu _{0}\right) $ shown in Fig. 2 has two singular points -- two
self-intersections for $\Omega \simeq 0.12$ and $\Omega \simeq 0.28$, see
Fig. 2. Green line corresponds to $G<G_{0}$, red line is the singular case, $%
G=G_{0}$, while blue line denotes the case $G>G_{0}$.

\begin{figure}[ht!]
\center 
\includegraphics[width=10.5cm, height=7cm]{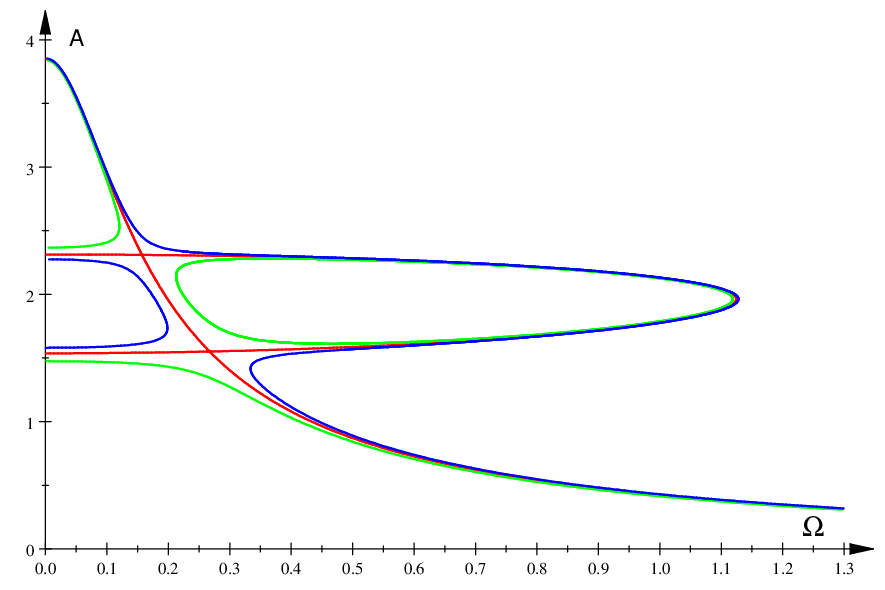}
\caption{The amplitude profiles $A(\Omega)$, $\protect\lambda =-0.118321$,$%
\protect\mu =3$ and $G=1.19$ (green), $G_0=1.214842$ (red), and $G=1.23$
(blue). }
\label{F2}
\end{figure}

Presence of these two singular points can be demonstrated by computing
bifurcations diagrams. 
\begin{figure}[h!]
\center 
\includegraphics[width=10.5cm, height=7cm]{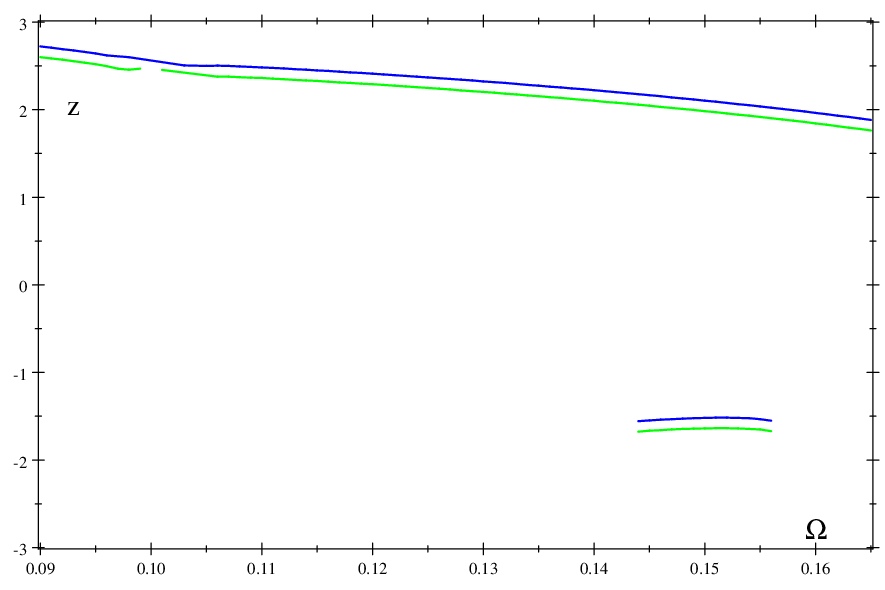}
\caption{Bifurcation diagram. $G=1.2418<G_0$, $\protect\mu =3$, $\protect%
\lambda =-0.118321$ (green), $G=1.2419>G_0$. $\protect\mu =3$, $\protect%
\lambda =-0.118321$ (blue). Blue line shifted vertically by $0.12$. }
\label{F3}
\end{figure}

In Fig. \ref{F3} the bifurcation diagram is shown for $G<G_{0}$, i.e. green
line in Fig. 2. For $\Omega <0.12$ only the upper green branch is stable.
Then, near $\Omega \simeq 0.12$ two upper branches disappear while the
lowest branch is still unstable. Then, for $\Omega \simeq 0.2$ there are
again three green branches in Fig. 2 -- the upper branch is stable while the
lowest branch is stable as well. In Fig. \ref{F3} we have $G>G_{0}$
corresponding to blue amplitude profile in Fig. 2. The upper branch is
stable all through and of two lower branches the lowest is stable. \vspace{%
0.2cm}

Induced dynamics near the second self-intersection is best demonstrated for $%
\mu =1$. For $\mu _{0}=1$ we get from Eqs. (\ref{IS}) solution $\lambda
_{0}=-0.149\,954$, $G_{0}=0.662\,017$ which fulfills $\Delta \left(
G_{0},\lambda _{0},\mu _{0}\right) =0$. Now we plot the corresponding
amplitude profile, i.e. the implicit function $A\left( \Omega \right) $
given by Eq. (\ref{AOmega}) for parameters values $\lambda =\lambda _{0}$, $%
\mu =\mu _{0}$, $G<G_{0}$, $G=G_{0}$, $G>G_{0}$. The amplitude function $%
A\left( \Omega ;G_{0},\lambda _{0},\mu _{0}\right) $ is shown in Fig. 4.

\vspace{0.5cm}

\begin{figure}[ht!]
\center 
\includegraphics[width=10.5cm, height=7cm]{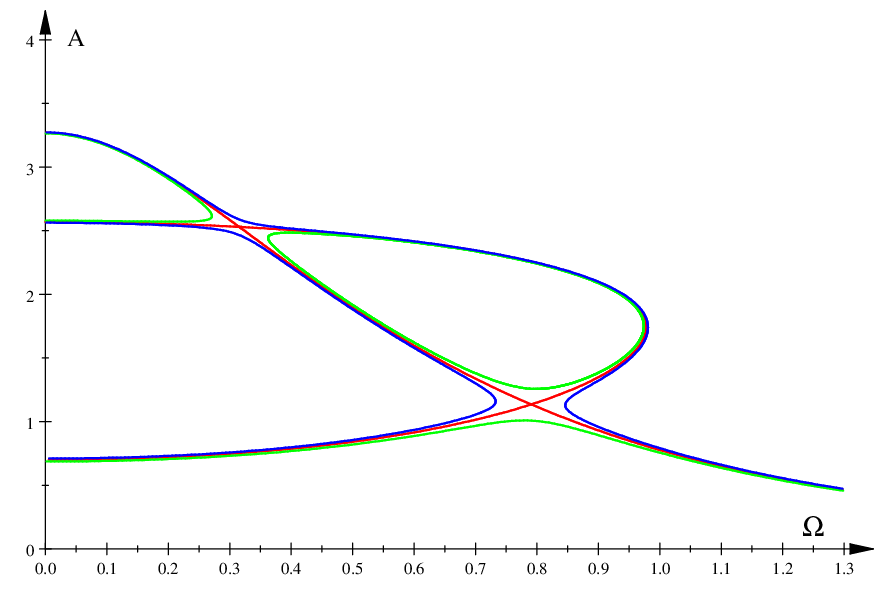}
\caption{ The amplitude profiles $A(\Omega)$, $\protect\lambda =-0.118321$, $%
\protect\mu =1$ and $G=0.65$ (green), $G=0.662017$ (red), and $G=0.67$
(blue). }
\label{F4}
\end{figure}
\vspace{0.3cm}

\begin{figure}[ht!]
\center 
\includegraphics[width=10.5cm, height=7cm]{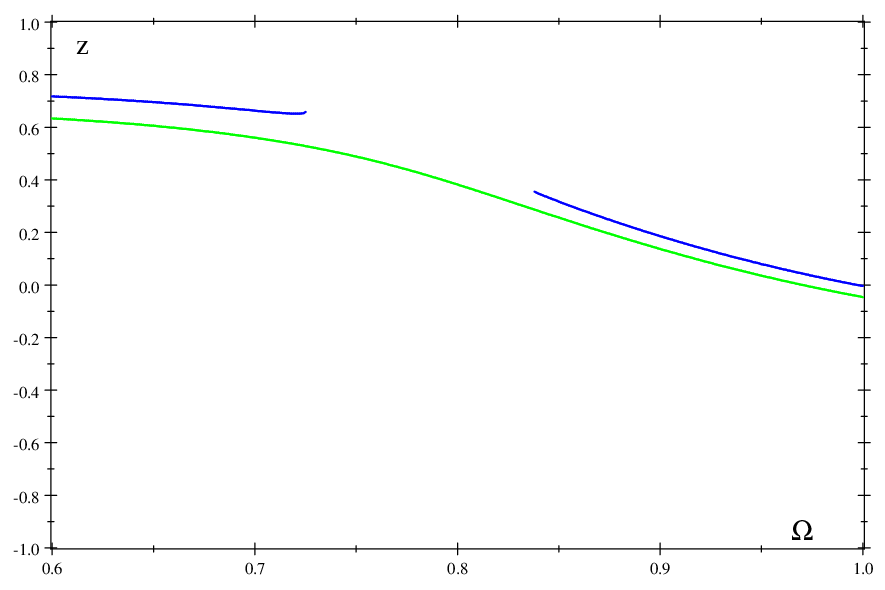}
\caption{Bifurcation diagram. $\protect\lambda =-0.118321$, $\protect\mu =1$%
, $G=0.65$ (green), and $G=0.67$ (blue). Blue line shifted vertically by $%
0.05$. }
\label{F5}
\end{figure}

Dynamics in the neighbourhood of the second self-intersection is shown in
Fig. \ref{F5}. It turns out that only lower branches are stable (for time
running backwards). Therefore, the blue branch has a gap, while the green
branch is continuous. \vspace{2.0cm}

\begin{figure}[ht!]
\center 
\includegraphics[width=10.5cm, height=7cm]{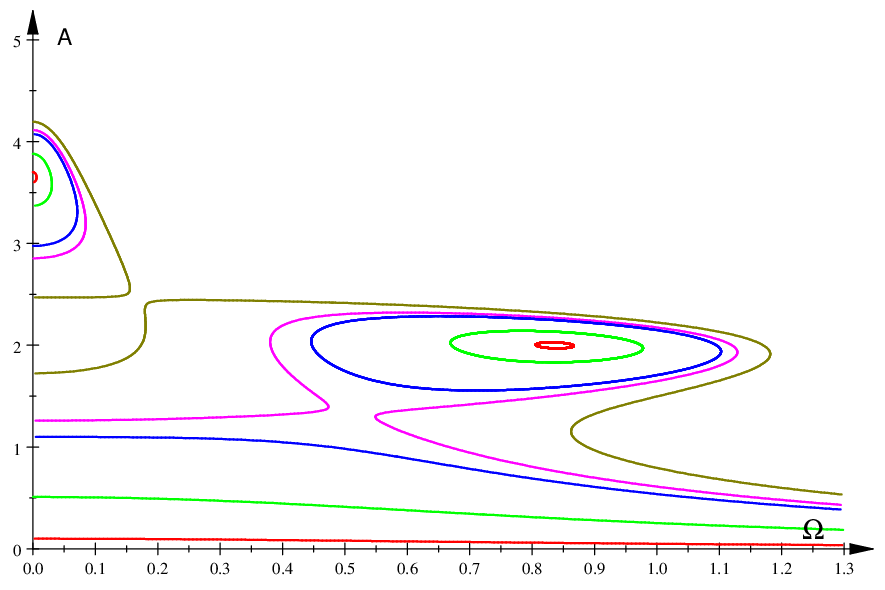}
\caption{ The amplitude profiles $A(\Omega)$, $\protect\mu = 2$, $\protect%
\lambda = -0.1$, $G=0.1$ (red), $G=0.5$ (green), $G=1$ (blue), $G=1.11$
(magenta), $G=1.34$ (brown). }
\label{F6}
\end{figure}

Finally, we compute the amplitude profile for $G \geq 0$, $\mu = 2$ and $%
\lambda = -0.1$. There are two singular points corresponding to $G=0$. The
amplitude profiles are shown in Fig. \ref{F6} for $G=0.1$, $G=0.5$, $G=1$, $%
G=1.11$, $G=1.34$.

\begin{figure}[ht!]
\center 
\includegraphics[width=10.5cm, height=7cm]{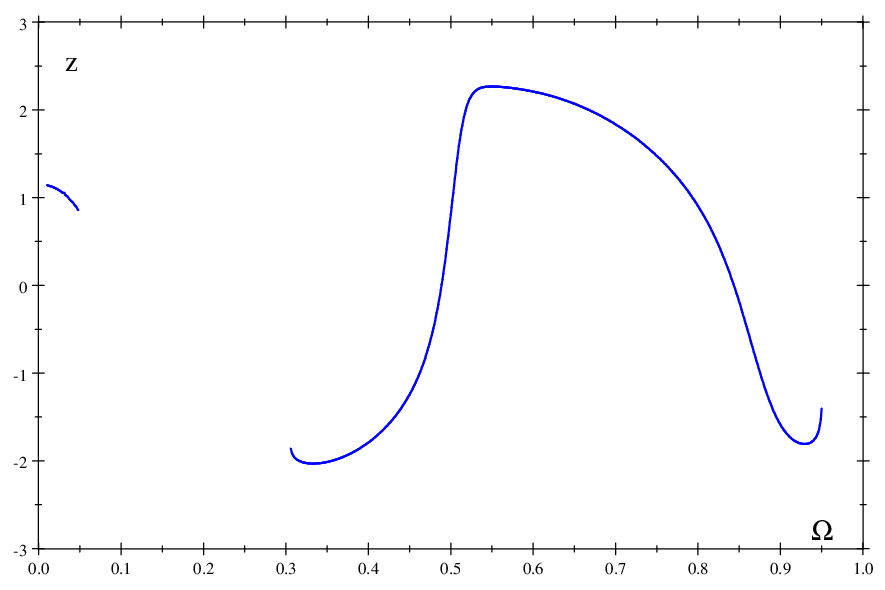}
\caption{Bifurcation diagram, $\protect\mu = 2$, $\protect\lambda = -0.1$, $%
G=1$ (cf. the blue line in Fig. \protect\ref{F6}). }
\label{F7}
\end{figure}

In Fig. \ref{F7} the corresponding stable branches are shown for $G=1$.

\section{Summary and discussion}

In this work we have investigated dynamics of the periodically forced
Duffing-van der Pol equation, continuing our study \cite{Kyziol2015a}.  We
were able to compute the bifurcation set -- the set of all parameters, such
that the amplitude profile has a singular point, see Eq. (\ref{Delta2}) and
Fig. 1. We have thus obtained a global view of singular points computing all
singular points in a novel way. Instead of solving the set of equations (\ref%
{SING1}) we have solved Eqs. (\ref{Sing1a}), (\ref{Sing1b}) for $Y$, $X$
obtaining Eqs. (\ref{Sol1}), (\ref{Sol2}) and demanded that Eq. (\ref{Sol1})
has multiple solutions. Note that Eq. (\ref{Sol2}) and Eq. (16) from Ref. 
\cite{Kyziol2015a} are identical. This suggests that the bifurcation set
contains all singular points since Eq. (16) determines the general solution
of Eqs. (\ref{SING1}) for $G \neq 0$. It seems that the method is general
and can be applied to other dynamical systems.

It is interesting that there are double singular points, i.e. there are two
singular points on the amplitude profile for some parameter values, see
Figs. 2, 4, 6. In Section \ref{Computations} the amplitude profiles and the
corresponding bifurcation diagrams have been computed to show metamorphoses
of dynamics near these singular points.

\appendix{} \label{App1}

\section{Coefficients of the polynomial $D$, Eq. (\protect\ref{Delta1})}

In this appendix coefficients of the polynomial (\ref{Delta1}) are listed
below. 
\begin{equation}
\left. 
\begin{array}{lll}
\mathbf{A}_{\mathbf{10}} & \mathbf{=\medskip } & 2^{8}\left( G+2+6\lambda
\right) ^{2}\left( G-2-6\lambda \right) ^{2}\medskip \\ 
\mathbf{A}_{\mathbf{8}} & = & b_{5}\lambda ^{5}+b_{4}\lambda
^{4}+b_{3}\lambda ^{3}+b_{2}\lambda ^{2}+b_{1}\lambda +b_{0} \\ 
b_{5} & = & 2^{12}3^{5} \\ 
b_{4} & = & -2^{12}3^{4}5 \\ 
b_{3} & = & -2^{8}3^{3}\left( 71G^{2}+2^{4}29\right) \\ 
b_{2} & = & -2^{8}3^{2}\left( 3\times 19G^{2}+2^{4}47\right) \\ 
b_{1} & = & 2^{6}3\left( -5G^{4}+2^{2}3^{2}7G^{2}-2^{11}\right) \\ 
b_{0} & = & 2^{6}\left( -5^{2}G^{4}+2^{2}7^{2}G^{2}-2^{9}\right) \medskip \\ 
\mathbf{A}_{\mathbf{6}} & = & c_{5}\lambda ^{5}+c_{4}\lambda
^{4}+c_{3}\lambda ^{3}+c_{2}\lambda ^{2}+c_{1}\lambda +c_{0} \\ 
c_{5} & = & -2^{15}3^{5} \\ 
c_{4} & = & 2^{15}3^{4}\left( G^{2}-1\right) \\ 
c_{3} & = & 2^{11}3^{3}\left( 157G^{2}+2^{4}5\right) \\ 
c_{2} & = & 2^{6}3^{2}\left( -5\times 7\times 11G^{4}+2^{2}23\times
83G^{2}+2^{9}11\right) \\ 
c_{1} & = & 2^{5}3\left( -5^{2}61G^{4}+2^{7}61G^{2}+2^{13}\right) \\ 
c_{0} & = & 3125G^{6}-24\,000G^{4}+43\,008G^{2}+65\,536\medskip \\ 
\mathbf{A}_{\mathbf{4}} & = & d_{5}\lambda ^{5}+d_{4}\lambda
^{4}+d_{3}\lambda ^{3}+d_{2}\lambda ^{2}+d_{1}\lambda +d_{0} \\ 
d_{5} & = & 2^{14}3^{5}\left( 7G^{2}+2^{2}\right) \\ 
d_{4} & = & 2^{13}3^{5}\left( 19G^{2}+2^{3}\right) \\ 
d_{3} & = & 2^{9}3^{4}\left( -3\times 5\times
11G^{4}+2^{3}3^{2}G^{2}+2^{7}\right) \\ 
d_{2} & = & 2^{9}3^{2}\left( -5^{2}31G^{4}-2^{2}5\times 47G^{2}+2^{7}\right)
\\ 
d_{1} & = & -2^{7}3^{3}G^{2}\left( 5^{3}G^{2}+2^{5}11\right) \\ 
d_{0} & = & -2^{12}3^{3}G^{2}\medskip \\ 
\mathbf{A}_{\mathbf{2}} & = & e_{6}\lambda ^{6}+e_{5}\lambda
^{5}+e_{4}\lambda ^{4}+e_{3}\lambda ^{3}+e_{2}\lambda ^{2} \\ 
e_{6} & = & 2^{16}3^{6}G^{2} \\ 
e_{5} & = & 2^{17}3^{5}G^{2} \\ 
e_{4} & = & -2^{15}3^{4}G^{2}\left( 3\times 5G^{2}+7\right) \\ 
e_{3} & = & -2^{14}3^{3}G^{2}\left( 5^{2}G^{2}+2\times 3^{2}\right) \\ 
e_{2} & = & -2^{12}3^{5}G^{2}\medskip \\ 
\mathbf{A}_{\mathbf{0}} & = & -2^{18}3^{5}G^{4}\lambda ^{5}%
\end{array}%
\right\}  \label{Coeff1}
\end{equation}

\section{Computational details}

\label{App2} Nonlinear polynomial equations were solved numerically using
the computational engine Maple from Scientific WorkPlace 5.5. All Figures
were plotted with the computational engine MuPAD from Scientific WorkPlace
5.5. Curves shown in bifurcation diagrams in Figs. \ref{F3}, \ref{F5}, \ref%
{F7} were computed running DYNAMICS, program written by Helena E. Nusse and
James A. Yorke.

\end{document}